# Visual Quality Enhancement in Optoacoustic Tomography using Active Contour Segmentation Priors

Subhamoy Mandal, *Graduate Student Member, IEEE*, Xosé Luís Deán-Ben, *Member, IEEE*, and Daniel Razansky*, *Member, IEEE*

*Abstract*— *Segmentation of biomedical images is essential for studying and characterizing anatomical structures as well as for detection and evaluation of tissue pathologies. Segmentation has been further shown to enhance the reconstruction performance in many tomographic imaging modalities by accounting for heterogeneities in the excitation field and tissue properties in the imaged region. This is particularly relevant in optoacoustic tomography, where discontinuities in the optical and acoustic tissue properties, if not properly accounted for, may result in deterioration of the imaging performance. Efficient segmentation of optoacoustic images is often hampered by the relatively low intrinsic contrast of large anatomical structures, which is further impaired by the limited angular coverage of some commonly employed tomographic imaging configurations. Herein, we analyze the performance of active contour models for boundary segmentation in cross-sectional optoacoustic tomography. The segmented mask is employed to construct a two compartment model for the acoustic and optical parameters of the imaged tissues, which is subsequently used to improve accuracy of the image reconstruction routines. The performance of the suggested segmentation and modeling approach are showcased in tissue-mimicking phantoms and small animal imaging experiments.*

## I. INTRODUCTION

THE unique capability of optoacoustics to render optical absorption contrast deep in optically scattering tissues with high spatial resolution has recently triggered tremendous development of imaging systems based on this technology[1],[2]. In particular, the multispectral optoacoustic tomography (MSOT) technique has emerged as a powerful tool to resolve spectrally-distinct absorbers by exciting the tissue with short laser pulses at multiple optical wavelengths[3], [4], [5]. In the common implementation of whole-body small animal imaging, MSOT can deliver cross-sectional images of a living mouse in real time by simultaneous collection of optoacoustic signals with a concave array of cylindrically focused transducers [6], [7]. Three-dimensional image data can be then achieved by sample translation along the elevation direction of the array. Similar detection geometries have been implemented in hand-held operation mode to enable clinical translation of the technology[8].

Efficient segmentation of optoacoustic images is essential for both anatomical characterization and enhancement of the image reconstruction performance. For instance, properly selected focus measures applied to the different parts of the reconstructed image can be used as a feedback mechanism to adjust the speed of sound (SoS) of the medium or other reconstruction parameters [9]. Image segmentation may enable identification of areas corresponding to different acoustic or optical properties. It has been previously shown that accounting for differences of SoS and attenuation between the imaged tissues and surrounding coupling medium (water) can improve imaging performance [10]. Researchers have used hybrid measurement approaches (passive ultrasound and optoacoustics) to account for acoustic heterogeneities, and accurately reconstruct optoacoustic images [11]. However, the methods render non-uniform resolution and require additional instrumentations. Born ratio-based optoacoustic image normalization has been suggested to correct for the heterogeneous light fluence distribution[12], [13] but the approach is not applicable for accounting for small SoS variations. Accounting for strong acoustic heterogeneities can mitigate image artifacts associated with acoustic reflections and scattering in biological tissues [14]. Additionally, identifying boundaries of regions with different optical properties is important for estimation of the light fluence distribution, which is essential for a quantified assessment of chromophore concentration in deep tissues [15].

Accurate image segmentation is generally considered a challenging task in image processing [16]. In spite of the

Manuscript received October 18, 2015, revised March 18, 2016 and April 04, 2016, accepted April 06, 2016. This work was supported in part by the ERC Starting Independent Researcher Grant under grant agreement ERC-2010-StG-260991. S Mandal acknowledges support from DAAD PhD Scholarship Award (A/11/75907) and IEEE Richard E. Merwin Scholarship. The authors acknowledge the support of NVIDIA Corporation with the hardware donation used for this research.

S Mandal is with the Department of Electrical and Computer Engineering, Technische Universität München, and Institute for Biological and Medical Imaging, Helmholtz Zentrum Munich; XL Dean Ben is with Institute for Biological and Medical Imaging, Helmholtz Zentrum Munich; and D Razansky is with the School of Medicine, Technische Universität München, and Institute for Biological and Medical Imaging, Helmholtz Zentrum Munich, Ingolstadter, Landstraße 1, 85764 Neuherberg, Germany. (*correspondence e-mail: s.mandal@ieee.org; dr@tum.de).

Supplementary video is added with the article, relevant code snippets are available at : http://www.mathworks.com/matlabcentral

difficulties, segmentation represents an efficient tool to improve inversion algorithms. For example, segmented CT priors have been used to improve fluorescence molecular tomography (FMT) reconstructions [17], and segmented magnetic resonance images have been used as priors for attenuation correction in positron emission tomography (PET)[18]. On the other hand, manual segmentation of images is useful to improve the quantification performance of optoacoustic tomographic reconstructions under heterogeneous illumination conditions [19] and in presence of strong acoustic reflections [20]. An automated segmentation process can then be integrated with advanced inversion algorithms to obtain more accurate reconstructions of the actual distribution of chromophores. In optoacoustic imaging, the task of segmentation is oftentimes exacerbated by the relatively low intrinsic contrast of large anatomical structures, having a much lower haemoglobin concentration than in major blood vessels, and is further impaired by the limited angular coverage of some commonly employed tomographic imaging configurations. Yet, proper segmentation of boundaries is essential to improve the quantitative performance and overall image quality of optoacoustic reconstructions. Even when considering relatively homogeneous tissue samples, accurate identification of the outer boundaries is essential for a proper assignment of acoustic and optical properties of the imaged region of interest and the background (coupling) medium [10].

This work investigates the applicability of active contour models [21], [22] for boundary segmentation in optoacoustic tomographic reconstructions. Specifically, the segmented boundary information is used to aid automated fitting of the SoS values in the imaged sample and the surrounding water. A reconstruction mask is further used for quantified mapping of the optical absorption coefficient by means of light fluence normalization. The performance of active contour segmentation for cross-sectional optoacoustic images and the associated benefits in image reconstruction are demonstrated in phantom and small animal imaging experiments.

## II. Modeling Methods

The active contour segmentation employed is described in this section along with the methods utilizing the segmentation mask to correct SoS heterogeneities and estimate the light fluence distribution within the imaged sample.

### A. Active Contour Models

Active contour, also referred to as snakes, is a deformable model widely used in image analysis applications given its flexibility and efficiency [22], [23]. Snakes are generally interactive in nature and demand user input or an initial guess, followed by movement of a curve towards the boundary of the object of interest. The mathematical formulation behind the snakes is an energy-minimizing spline, with an associated 'snake energy' as a cost function. The spline is guided by external constrain forces and influenced by image forces which oblige it to converge towards dominant features like lines or edges [21]. In the current study, the classical snake model is used. It involves a controlled continuity spline representing a generalization of Tikhonov stabilization, which can be visualized as a regularization problem [24], [25]. The governing forces for curve evolution are: (a) image forces $E_{image}$ that push snakes towards the edge feature and contours, (b) the internal energy $E_{int}$ enforcing a piecewise smoothness constraint, and (c) the external constrained forces $E_c$ that put the snakes closer to appropriate local minima. Thus the total energy of the image can be expressed as

$$E_{snakes}(s) = E_{image}(s) + E_{int}(s) + E_c(s) \quad ..(1)$$

where $s$ encodes the parametric snake representation. The optimum snake is then obtained as

$$s_{opt} = \ \arg_s \min E_{snakes}(s) \quad ....(2)$$

It should be noted that $E_{image}$ is purely data driven, whereas $E_c$ is defined by user interactions. As $E_{image}$ is influenced by the nature of the image, it needs to be carefully calibrated. In our case, calibration is achieved by using the energy functional as proposed by Kass et al. [13] i.e.,

$$E_{image} = \ w_{line} \ E_{line} + w_{edge} \ E_{edge} + w_{term} \ E_{term} \quad ..(3)$$

where $E_{line}$, $E_{edge}$ and $E_{term}$ are the energy functionals and w represents the corresponding weights. A more detailed mathematical explanation is available as appendix [26]–[28].

Further, a morphological image processing sub-unit was added to the image segmentation workflow to generate smoother boundaries [29]. The morphological processes utilized a disc shaped structuring element with 2 pixel diameter. Closing operations were carried out on the segmented (binary) image mask to plug any spurious holes or irregular edge inundations that might be present in the mask due to limited view problems.

Deformable models and active contour segmentation is in itself a highly investigated area in image analysis and several advanced methods viz. active contour without edges [26], localized region based active contours [27], geodesic methods [30] etc. exist. However, we limit our current investigation to the classical model postulated by Kass et. al. [21] and design an efficient workflow with the existing algorithm to achieve the goal of segmenting MSOT images and using this information as prior for mapping optical and acoustic inhomogeneity, ultimately leading to better imaging performance.

### B. Correction of light attenuation using segmentation prior

In MSOT, the reconstructed images for each wavelength represent a map of the spatially varying optical energy being absorbed by the tissue. Assuming a uniform Grueneisen parameter in soft biological tissues, the absorbed energy is proportional to the product of the absorption coefficient $\mu_a(\vec{r})$ and the light intensity $U(\vec{r})$ at a given voxel. Obtaining accurate maps of concentration of individual chromophores implies on the one hand spectrally or temporally unmixing the distribution of the chromophore of interest from other substances contributing to $\mu_a(\vec{r})$, and on the other hand normalizing the optoacoustic images with the excitation light fluence distribution. Several mathematical procedures have been suggested for light fluence normalization [31], including optical propagation models, iterative algorithms based on fixed-point methods [32], [33], frequency decomposition of optoacoustic images, logarithmic unmixing of multispectral datasets [34] or estimation of decay rates of photoswitchable probes [35]. Light fluence can be directly estimated from the optoacoustic images [12], but the image need to be properly segmented to accurately assign the optical properties to different tissues. Here, we propose a diffusion-theory-based light propagation model that uses the segmented boundaries of the object as a method to extract quantified information from optoacoustic images. The photon diffusion equation [36], [37] is assumed to model the light intensity in scattering tissues, i.e.,

$$S_0 = -\nabla.[D(\vec{r}).\nabla U(\vec{r})] + \mu_a(\vec{r})U(\vec{r}) \dots (4)$$

where $D = \frac{1}{3}(\mu'_s + \mu_a)$ is the spatially dependent diffusion coefficient of the medium and $\mu'_s$ is the reduced scattering coefficient, $S_0$ being the source term. Eq. (4) neglects transient effects, which applies for typical pulse durations of 10 ns (or equivalently 3 m length) used for optoacoustic excitation. We assumed a constant intensity illumination ($S_0$) on the boundary enclosing the object volume, and the problem was treated as a two-dimensional problem for simplicity. The interface between the scattering and non-scattering medium can be modelled e.g. using the Robin boundary condition [38],

$$U(\vec{r}) + 2D(\vec{r})\hat{n}.\vec{\nabla}U(\vec{r}) = 0, \vec{r} \in \partial\Omega \dots(5)$$

where $\partial\Omega$ is the boundary of the object and $\hat{n}$ is the unit vector normal to the boundary pointing outwards. The solution to eq. 5 is obtained numerically using a finite volume method (FVM) solution approach based on the Deal II Framework [39]. In the given framework the elements of the mesh have a cubical base shape which is deformed and stretched to match the shape of the imaged object. The value of the fluence is thereafter calculated for the nodes and interpolated over the elements using a pre-defined interpolation function to reduce the computational complexity. As opposed to the earlier approaches that employed a geometrical approximation of the thresholded image to generate the FVM mesh and mapping the nodes [15], [39], we herein used the precisely segmented boundary to provide a more accurate basis for generating the mesh.

The initial images used to test the performance of the suggested fluence correction approach were reconstructed with the interpolated-matrix-model inversion IMMI method [40]. IMMI retains the quantitative nature of MSOT images by using exact numerical model-based reconstruction and reducing artifacts, and thus is preferred for accurately mapping the distribution of chromophores. A more detailed explanation of the IMMI algorithm is added in the appendix section [40]–[42].

### C. Multiple speed of sound fitting

In a typical MSOT imaging domain, the coupling medium and the imaged object often have different SoS, which may impair the image reconstruction quality [9], [10]. Different approaches have been so far developed to alleviate this problem, e.g. by means of the SoS mapping or generalized Radon transform models assuming geometrical acoustics approximations. Accordingly, here we use the segmented image mask to fit multiple SoS values to the different regions in the image, thus providing an alternative method to address the issue of acoustic heterogeneities without increasing the mathematical or hardware complexity of the imaging problem. To evaluate the performance of the proposed methodology, a two compartment model is considered consisting of the coupling medium outside the imaged object (the background) assigned with a given SoS ($c_b$) and the region inside the imaged sample assigned with a different SoS ($c_0$). In this formulation, the method is applicable to any reconstruction algorithm capable of accounting for a known distribution of the SoS, e.g. the filtered back-projection (FBP) algorithm [43], the interpolated matrix IMMI method [40] or the time-reversal approach [44]. Herein, the FBP algorithm was used to automate the workflow in conjunction with automated SoS calibration. The automated SoS calibration for the image fitted with two values of the SoS was carried out with a normalized variance of the image gradient magnitude using Sobel operator (Sobel+Var) and anisotropic diffusion using consistent gradient operator (Ad-CG), as recently reported by Mandal et al [9], [45].

The FBP method is based on a delay-and-sum approach and for a finite number of measuring locations, where the optical energy deposition $f(x'_j, y'_j)$ at a given pixel of the region of interest (ROI) is calculated as

$$f(x'_j, y'_j) = \sum_i s(x_i, y_i, t_{ij}) \dots (6)$$

where $(x_i, y_i)$ is the $i^{th}$ measuring location and $t_{ij} = |(x'_j, y'_j) - (x_i, y_i)|/c$, being $c$ the SoS when a uniform SoS is considered. $s(x_i, y_i, t_{ij})$ represents the function to be back-projected, i.e. filtered pressure. Eq. (6) is expressed in arbitrary units, where the constant terms accounting for unit conversion factors are omitted for simplicity. As mentioned earlier, we fit two SoS values ($c_b$, $c_0$) for the background and object respectively. In this case, $t_{ij}$ in Eq. (6) is estimated as the time of flight (ToF) from the center of the transducer to a given voxel calculated as

$$ToF = \frac{d_c}{c_0} + (d - d_c)/C_b \quad \dots (7)$$

where d is the distance from the detector to the voxel considered and $d_c$ is the fraction of d located within the tissue.

We can employ the IMMI algorithm for multiple SoS mapping, however the method is computationally expensive for SoS calibration purposes. Thus, we choose FBP methods for automated SoS calibration and multiple SoS fitting. The details of the SoS calibration methods and the rational for favoring the faster (but often less accurate) FBP method has been illustrated in [9].

## III. EXPERIMENTAL DESCRIPTION

### A. Experimental Setup

The experiments were conducted using a commercial small animal MSOT scanner (MSOT256-TF, iTheraMedical GmbH, Munich, Germany). The system is designed to acquire cross-sectional optoacoustic images using its custom-made 256-element concave array of cylindrically-focused piezocomposite transducers (5 MHz central frequency) with a radius of curvature of 40 mm and approximately 270° angular coverage. The optoacoustic signals generated with each laser pulse are simultaneously sampled at 40 Mega samples per second by parallel analog to digital converters. Light excitation is provided with the output of a laser beam from a wavelength-tunable optical parametric oscillator (OPO)-based laser, which is shaped to attain ring-type uniform illumination on the surface of the phantoms by means of a custom-made fiber bundle [6]. The customized Nd:Yag pump OPO laser (Innolas Laser GmbH, Germany) is tunable between 680-900 nm with pulse width of ~10 ns (10 Hz repetition rate) and ~90 mJ per-pulse energy at the laser's output .

### B. Tissue Mimicking Phantoms

Two types of tissue mimicking phantoms were developed for testing purposes. The first phantom was designed to mimic a small increment in the SoS. This was achieved with a mixture of agar solution (approximately 2/3 by volume) and glycerine (approximately 1/3 by volume), as described in [10], [42]. As glycerine is hydrophilic, it readily dissolves in the water-based agar solution. Given the fact that the SoS in glycerine is 1920 m/s and the SoS in agar gel is approximately 1500 m/s, the expected SoS in the mixture is around 1640 m/s, i.e., ~10% increase. Black polyethylene microparticles with a diameter of approximately 200 μm (Cospheric LLC, Santa Barbara, CA) were then embedded in the imaging plane to assess the spatial resolution rendered. The second phantom was built to render a light fluence attenuation representative of average soft tissues. For this, black India ink and Intralipid were added to the agar solution in order to attain an optical absorption coefficient $\mu_a = 0.2$ cm$^{-1}$ and a reduced scattering coefficient $\mu_s' = 10$ cm$^{-1}$. Two tubular insertions of a more concentrated India ink having an optical absorption coefficient of $\mu_a = 1$ cm$^{-1}$ were embedded in the phantom at different depths, one tubing being located close to the periphery and the other approximately at the center of the phantom. The phantoms were made in irregular shapes so that the efficiency of segmentation algorithm can be properly assessed. All phantoms were finally embedded in cylindrical blocks of agar with a diameter of 24 mm for easier handling.

### C. Small animals imaging

The segmentation performance was evaluated with *in-vivo* mouse images. For this, the animal handling protocols were scrupulously followed under supervision of trained personnel and the imaging experiments were performed in full conformity with institutional guidelines and with approval from the Government of Upper Bavaria. The mice were sedated with Isoflurane and immersed in water bath kept at 34°C using a specialized mouse holder (iThera Medical GmbH, Munich, Germany). Altogether, 30 datasets were acquired from 12 different mice for three different representative anatomical regions, namely (a) brain, (b) liver and (c) kidney/spleen. In addition, two polyethylene tubings containing Indocyanine green (ICG) $\mu_a = 1.9$ cm$^{-1}$ at $\lambda = 800$nm were inserted at different depths (peripheral T1 and deeply embedded T2) in a CD1 mouse *post mortem* in order to quantitatively validate the suggested SoS and light fluence correction methods in real tissues [12]. The mouse was then imaged at different positions along the torso using 10 different wavelengths ($\lambda$) ranging between 680 and 900 nm. From the multispectral data, the ICG distribution was spectrally unmixed from background tissue components using a semi-automated blind unmixing algorithm termed vertex component analysis[46].

## IV. RESULTS

The phantom results are presented in Fig. 1. Fig 1a shows the tomographic reconstruction obtained when a uniform value of the SoS is assumed for both the phantom and the surrounding coupling medium. A single value of the SoS (c=1535m/s) was fitted based on focusing metrics, which visually yields a reasonable reconstruction. However, a closer observation reveals that the difference of SoS between background and phantom has inevitably led to degradation of image quality. Specifically, a zoom in on a selected microsphere reveals that some microspheres in the imaged plane have not been accurately reconstructed due to acoustic mismatch and wrong assignment of the speed of sound distribution. Additional artifacts appear at the edge of the phantom, as indicated by the white arrow. Active contour segmentation (morphological processing as explained in section 2.A) was subsequently applied to the tomographic reconstruction in Fig. 1(a), yielding the mask displayed in Fig. 1(b). The mask allows differentiating between the imaged object and the background region, so that different values of the SoS can be assigned. In particular, the values of the SoS for the sample ($c_0$=1565m/s) and the background ($c_b$=1515m/s) were estimated based on autofocusing metrics. The corrected tomographic reconstruction obtained by considering the mask in Fig. 1(b) and the fitted values of the SoS is showcased in Fig. 1(c). The magnification of the same area as in Fig. 1(a) shows a sharper reconstruction of the microsphere. Furthermore, the model using multiple SoS renders a sharper appearance of the edges (marked with a white arrow).

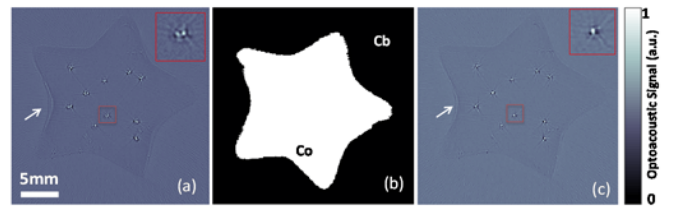

Fig. 1. A single value of the speed of sound (SoS = 1535) assumed homogeneous is applied to obtain (a), the segmentation mask (b) is extracted using active contour method and morphological processing. Two different SoS -1525 ($c_b$) for background region and 1565 ($c_0$) for inside object boundary were fitted to obtained an improved reconstruction (c). Zoom in (region marked with red boxes) of reconstructed microsphere(s) is included in the insert and reconstructed edges are marked with arrows.

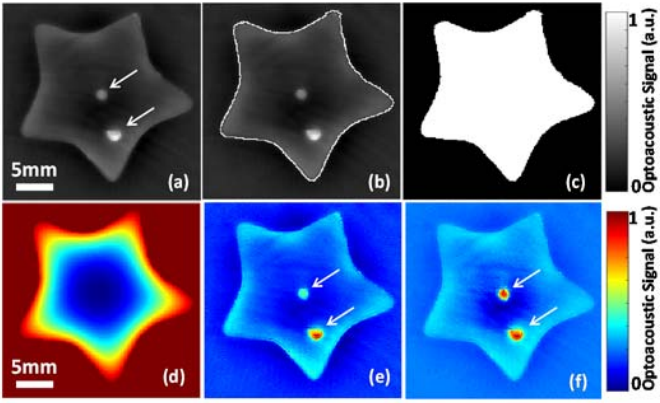

Fig. 2. Model-based (IMMI) reconstruction without optical fluence correction shows an erroneous absorption distribution (a). The reconstructed image is segmented using the active contour method (b) and the binary segmentation mask is extracted (c). The fluence model (d) is generated using the mask (b) as prior by application of the FVM method. Two tubular insertions of India ink having equal optical density but placed at different depths in a turbid medium were imaged (marked by white arrows). The fluence corrected image (f) using the proposed method shows better visual saliency than uncorrected image (e).

The results of the segmentation-assisted optical fluence normalization are illustrated in Fig 2. First, the optoacoustic image of the tissue mimicking phantoms with two identical tubular insertions was obtained using the IMMI method, as shown in Fig. 2(a). It can be readily seen that the two insertions (marked with white arrows) appear with substantially different intensities in the image. Indeed, optical attenuation - contributed by absorption and scattering − in the tissue mimicking phantom leads to a lower signal intensity generated for deeper-seated objects. Active contours were then used to segment the boundary of the sample, as shown in Fig. 2(b). Based on the segmented mask in Fig. 2(c), the light fluence distribution was obtained with a FVM based simulation assuming uniform light distribution on the boundary of the object and uniform optical properties (Fig. 2(d)). Normalization of the optoacoustic image with the estimated light fluence field yields the image displayed in Fig.

2(f) (Fig. 2(e) is the reference image without normalization), in which the two embedded tubings show similar signal intensity as would be expected from absorbers comprising the same concentration of the chromophore.

The applicability of active contour segmentation for realistic in-vivo small animal datasets reconstructed with the IMMI method is showcased in Fig. 3. The optimum values of the terms in Eq. 3 for in-vivo mouse imaging were

TABLE I
ACTIVE CONTOUR SEGMENTATION PARAMETERS

| Regions | Filter Coef | # Iter | $W_{LINE}$ | $W_{EDGE}$ | $W_{TERM}$ | DC |
|---------|-------------|--------|------------|------------|------------|-----|
| *Brain* | 0.5 | 550 | 0.5 ±0.1 | 0.6 ±0.2 | 0.7 | 0.944 |
| *Liver* | 1.0 | 250 | 0.3± 0.1 | 0.5± 0.05 | 0.7 | 0.957 |
| *Kidney/Spleen* | 0.7 | 280 | 0.3 ± 0.1 | 0.5 ±0.1 | 0.7 | 0.953 |

Filter Coef. denotes the values used for Gaussian kernel before initiation of the snakes. Iter denotes the number of iterations applied for each region. W represents the values for energy functions as employed for the line, edge and term measures respectively. DC gives the value for Dice Coefficients, which is a measure of segmentation accuracy.

heuristically determined by computing the segmentation with different values and assessing the segmentation performances over multiple datasets acquired from the brain, liver and kidney/spleen regions. These regions have image properties and morphologies that are very distinct from each other, thus the properties of the snakes were recalibrated for each region individually. We used 10 datasets for each of the regions and used their averaged values to determine the mean parameters, as listed in Table 1.

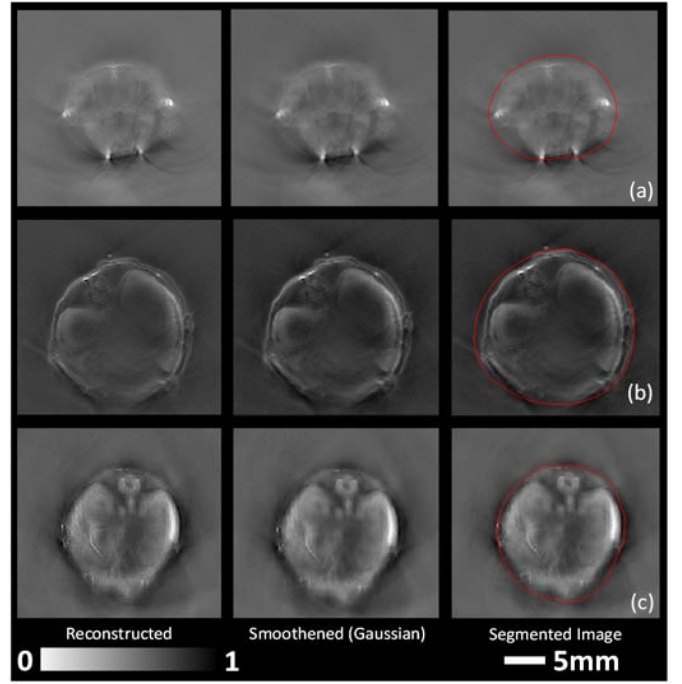

Fig. 3. Tomographic optoacoustic reconstructions of the brain (a), liver (b) and kidney/spleen (c) regions of mice in vivo. The original reconstructed images obtained with model-based inversion are shown in the first column. The second column displays the smoothened images after Gaussian filtering. The segmented images using active contour (snakes) with the optimum parameters are showcased in the third column. Movie displaying the segmentation process with step wise implementations and segmentation performances with different number of iterations is available as supplementary material.

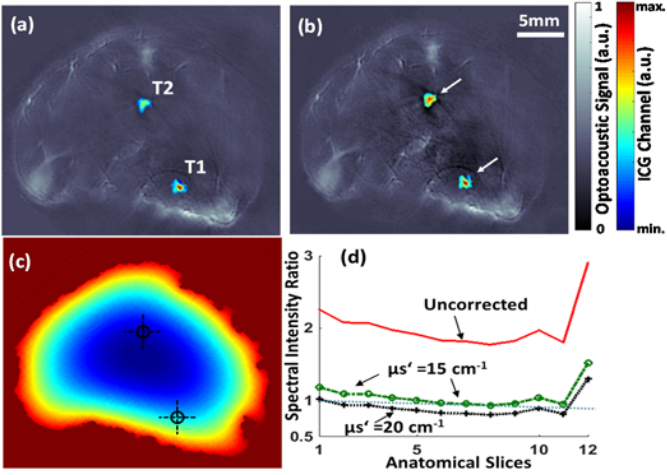

Fig. 4. Post-mortem imaging of mice with ICG insertions $\mu_a$ = 1.9 cm⁻¹ at $\lambda$ = 800nm at two different depths (T1 and T2): The optoacoustic anatomical reference image and the unmixed images corresponding to the ICG distributions for the uncorrected image (a) and fluence corrected image (b) are displayed. Both images were obtained post unmixing using 10 different wavelengths. The fluence map generated using the FVM method using the segmented prior is shown in (c), crosshairs indicates positions of insertions. The relative contrast ratios for fluence uncorrected and corrected images (12 anatomical slices) for two different tissue absorption properties are plotted in (d), clearly demonstrating a significant improvement (~1.1691) after application of fluence correction using the proposed segmented prior based FVM method.

Further, the segmentation performance was tested using both heuristically chosen initial starting contour points as well as by using geometrical shape priors for initiating the contour evolution automatically. The experimental results also revealed that the minimum number of contour points that need to be defined for curve initiation is 9, so that this configuration has been used through the rest of the test samples. For better segmentation performance, a Gaussian filtering was further applied to the original images before initiation of the curve evolution. The filter coefficients used for different in-vivo datasets are shown in Table 1.

The results show high variability of the parameter sets in the brain region, which demand more iteration steps (slower convergence) and yields lower segmentation accuracy. The Dice Coefficient (DC) was then used for characterizing the segmentation accuracy vis-à-vis with user feedback. DC provides us with a statistical measure for comparing similarity of two samples, and its cut-off limits for good segmentation are empirically decided by studying the ecologic association between species in nature [47]. For this, four independent volunteers performed manual segmentation of the reconstructed images. The current results yielded a DC ≥ 0.9, whereas DC > 0.7 is generally considered as good segmentation. In this way, a satisfactory performance of active contour models for MSOT image segmentation was confirmed.

The results of the post-mortem imaging studies are displayed in Figs. 4 and 5. Fig. 4 shows an example of a cross-sectional image being corrected for the light fluence attenuation effects. The ICG tubings and the average tissues were spectrally unmixed into two components and superimposed with different colormaps after applying the fluence correction for better visualization. It clearly shows that the relative signal intensity increases for the deep-seated tubing (T2) after light fluence normalization (Fig. 4(b)) as compared to the non-normalized image (Fig. 4(a)). To quantitatively evaluate the performance of the method, two-dimensional regions of interest surrounding the tubing locations were considered and the relative intensity values were compared before and after light fluence normalization for the different cross-sections (Fig. 4(d)). Uniform optical properties were assumed within the sample according to reported average values for biological tissues [48]. The absorption coefficient was taken as $\mu_a$ =0.29cm⁻¹, while the light fluence distribution was estimated for two different values of the reduced scattering coefficient, namely $\mu_s'$ =15cm⁻¹ and $\mu_s'$ =20cm⁻¹ [32], [49]. In both cases, the calculated intensity ratios were close to 1 for the normalized cross-sections, which demonstrates that the proposed method can offer good performance with segmented image priors. Thus, in spite of considering a simplified model for representing heterogeneous tissues, an improvement in the quantitative performance of MSOT is achieved.

The post-mortem acquisitions were also used for evaluating the algorithm performance in the case of low contrast ex-vivo images that may yield less accurate segmentation using active contours. Here we first reconstructed an image using the backprojection algorithm, whereas focusing matrices with temperature priors were used to determine the most suitable SoS. An initial segmented mask was then obtained by segmenting the reconstructed image, which was morphologically processed to provide a smoother mask. Finally, the obtained mask was employed to build the two-compartment model of SoS. The SoS of the background coupling medium (water bath) was determined using its measured temperature ($c_0$=1515m/s at 34°C). For computing the SoS of the object, we reconstructed a stack of 100 images with SoS spaced 5 m/s and a fixed background SoS. Selected autofocusing metrics (as mentioned in section 2.C) were executed to determine the calibrated SoS ($c_b$=1555 m/s). The reconstructed image obtained with the two compartment model is shown in Fig 5(b). The autofocusing metrics showed good convergence in the presence of strong absorbers in Fig. 5(d), and the line profiles displayed in Fig 5(c) indicate significant improvements with respect to images reconstructed with a uniform SoS. Indeed, not only are the tubings better characterized but also the internal structures are observed to be marginally more defined, e.g. spinal region, when using the two compartment SoS approach. Similar to the light fluence normalization measurements, segmentation of internal areas having different values of the SoS may further improve the resolution and contrast of the reconstructed images obtained by accounting for heterogeneous acoustic properties of tissues.

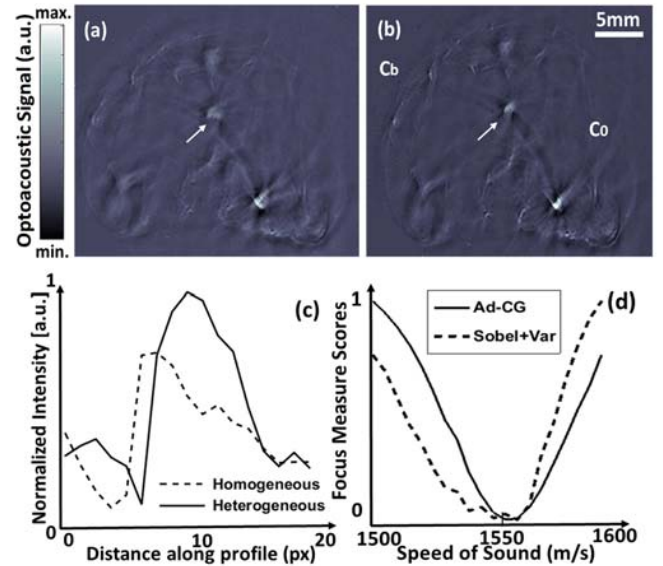

Fig. 5. Post-mortem images of mice with ICG insertions (equal diameter) at two different depths reconstructed with (a) single speed of sound (1520) and (b) multiple (differential) speeds of sound for background an object ($c_b$ = 1515 and $c_0$ = 1555 m/s). The line profiles of the marked insertion along the direction of the arrow when reconstructed with homogenous and heterogeneous SoS are shown in figure (c). The automated speed of sound calibration curves (Sobel +Var and Ad-CG methods) for heterogeneous SoS fitting are shown in (d).

## V. Discussion and Conclusion

In this article we have successfully demonstrated the applicability of active contour (snakes) segmentation in cross-sectional (two-dimensional) tomographic optoacoustic imaging. We have shown that using the segmentation results as prior information during the image reconstruction procedure can significantly improve the imaging performance. The performance of active contour models was outlined herein for whole-body segmentation of mice using three representative anatomical regions. Considering that the recently developed hand-held probes for optoacoustic clinical translation employ the same acquisition geometry, the methods presented in this work can further be applied in the development of diagnostic tools for translational imaging in human subjects [50]. The optimum values of the parameters used in the segmentation procedure such as the weighting factors to estimate the image forces or the number and location of the initial contour points were determined heuristically. Specifically, these values were computed over multiple datasets and the segmentation performance was subsequently evaluated. Thereby, these parameters may need to be recalibrated for other tomographic configurations or other biological samples. Further, it is possible to develop strategies to automate the seed point detection through the use of parametric curve fitting models, given a computational overhead [51]. The fitting models can further reduce dependency on human feedback, but do not influence the final segmentation performance.

Acoustic inversion in optoacoustic tomography is commonly conditioned by the differences between the acoustic propagation properties of the imaged tissue versus the surrounding coupling medium. Typically, a uniform non-attenuating medium comprising both the sample and water is assumed for reconstructions, which may lead to an inadequate imaging performance since the average acoustic properties in soft biological tissues are generally different from water. Thereby, identifying regions with different speed of sound or acoustic attenuation or areas with strong acoustic scattering or reflections is essential for optimizing the reconstruction performance. A two compartment model assuming an acoustically homogeneous tissue can then be used as a second order approximation to improve accuracy of the tomographic reconstructions. We have shown that by considering a different SoS in the sample, one can improve the spatial resolution performance when imaging through real heterogeneous tissues. Clearly, a similar procedure can be used to segment the imaged medium into three or more compartments in order to further enhance the imaging performance.

In a similar manner, optical propagation is very different in a transparent coupling medium as compared to real biological tissues, where photons undergo strong absorption and scattering. Thereby, a two compartment model can also be used as a first order approximation to account for the significant effect of light fluence attenuation, and segmentation of the optoacoustic images is then again essential for this purpose. Normalization with the light fluence distribution is needed for quantifying the concentration of specific optical biomarkers. We have shown that a numerical model assuming homogenous optical properties within properly segmented boundary can already be sufficiently accurate for normalization purposes both in phantom experiments with controlled absorption and scattering, as well as in real biological samples. A more accurate estimation of the light fluence within the sample implies detailed knowledge of the spatial distribution of its optical properties, which is very challenging to obtain in living organisms. It was previously shown that iterative light normalization methods may render good performance in numerical simulations [32], [49] but can turn unstable in experimental studies if the scattering coefficient or object boundaries are not accurately estimated [15], [19]. Proper segmentation of the outer boundary of the sample or internal regions having strong changes in optical properties then becomes essential for accurate light fluence normalization in optoacoustic tomography.

In conclusion, optoacoustic image segmentation and analysis is a nascent and emergent area of investigation that holds potential for advanced applications including, but not limited to organ segmentation, diagnostic imaging [52] and pharmacokinetic studies [53]. The demonstrated good performance of active contour models for optoacoustic segmentation of real tissues, and the feasibility to account for optical and acoustic discontinuities in the imaged region, anticipate the general applicability of the suggested approach for enhancing the resolution and image quality of tomographic optoacoustic reconstructions.

## APPENDIX

### Active Contours:

In our study we implemented the point snake model, proposed by Kass et al [21]. which employs an elementary representation of discrete curves, and satisfy a n-neighbor connectivity [22]. As mentioned in II. A, themodel consists in a controlled continuity spline influenced by image forces as well as external constrained forces. Given that the controlled continuity spline is a generalization of the Tikhonov stabilizer [24], Kass et al. suggested to treat it as a *regularization* problem [25].

Geometrically, snakes are contours embedded in an image plane $(x, y) \in \Re^2$. The position of a snake can be parametrically represented as $v(s) = (x(s), y(s))$, where $x$ and $y$ are the coordinate functions and $s \in [0, 1]$ is the domain. The energy functional of the contour is given by

$$E^*_{snake} = \int_0^1 E_{snake}(v(s))ds$$
$$= \int_0^1 E_{int}(v(s)) + \int_0^1 E_{image}(v(s)) + \int_0^1 E_c(v(s))ds \quad ..(8)$$

where $E_{int}$ is the internal energy (piecewise smoothness constraint), $E_{image}$ represents the image forces, and $E_c$ gives the external constrained forces (pushes snakes to desired local minima). Eq. (1) in section II.A represents a simplified version.

Deconstructing each component, the internal energy can be depicted as

$$E_{int} = (\alpha(s)|V_s(s)|^2 + \beta(s)|V_{ss}(s)|^2)/2 \quad ..(9)$$

with the first and second order control terms, $\alpha(s)$ and $\beta(s)$ respectively, controlling the nature and behavior of the curve. Specifically, $\alpha(s)$ controls the tension, and $\beta(s)$ controls the rigidity in an image [23]. For discretizing the energy formulations Eq 1.2 can be rewritten by using vector notation $V_i = (x_i, y_i) = (x(ih), y(ih))$ as

$$E_{int} = \alpha_i |V_i - V_{i-1}|^2 /2h^2 + \beta_i |V_{i-1} - 2V_i - V_{i+1}|^2 /2h^4 \quad ..(10)$$

The corresponding Euler equations obtained can be represented in matrix format are

$$\mathbf{Ax} + \mathbf{f_x(x, y)} = \mathbf{0}$$
$$\mathbf{Ay} + \mathbf{f_y(x, y)} = \mathbf{0} \quad ..(11)$$

Solving the equations in a matrix inversion, we obtain

$$\mathbf{x_t} = (\mathbf{A} + \gamma\mathbf{I})^{-1}(\gamma\mathbf{x}_{t-1} - \mathbf{f_x}(x_{t-1}, y_{t-1}))$$
$$\mathbf{y_t} = (\mathbf{A} + \gamma\mathbf{I})^{-1}(\gamma\mathbf{x}_{t-1} - \mathbf{f_x}(x_{t-1}, y_{t-1})) \quad ..(12)$$

where $\mathbf{A}$ and $\mathbf{A} + \gamma\mathbf{I}$ are pentadiagonal banded matrices, and $t$ represents the time steps. Theinverse matrix is calculated using LU (lower-upper) decomposition, being the computational complexity $O(n)$.

The image energy, as mentioned, can be expressed as a combination of weighted energy terms as

$$E_{image} = w_{line} E_{line} + w_{edge} E_{edge} + w_{term} E_{term} \quad ..(13)$$

The line functional is represented by image intensity $E_{line} = I(x, y)$, and the weight $w_{line}$ is determined by the pixel intensity towards which it is attracted.

The edge functional is given by the filtered image intensity gradient $E_{edge} = -(Filt * \nabla^2 I)^2$, where $Filt = G_\sigma$. We introduce the filtering term to avoid the spline from getting struck in a local minimum. Mandal et al [9] has shown the applicability of scale-space processing for optoacoustic images, and an anisotropic diffusion (Modified Perona-Malik diffusion) with Gaussian kernel can also be employed,

$$\frac{\partial I}{\partial t} = div(c(|DG_\sigma * I|)\nabla I)$$

where D is the diffusion tensor and c is the diffusion coefficient [45], [51].

The terminal functional plays of role in finding terminations of line segments and corners and operates on a smoothened image $I' = (Filt * I)$. Assuming the gradient angle $\theta = \tan^{-1}(I'_y/I'_x)$, we can write the curvature of level contours in $I'(x, y)$ as

$$E_{term} = \frac{\partial \theta}{\partial \mathbf{n}_\perp}$$
$$= \frac{\partial^2 I'/\partial \mathbf{n}^2_\perp}{\partial I'/\partial \mathbf{n}}$$
$$= \frac{I'_{yy}I'^2_x - 2I'_{xy}I'_x I'_y + I'_{xx}I'^2_y}{(I'^2_x - I'^2_y)^{3/2}} \quad ..(14)$$

where $\mathbf{n} = (\cos\theta, \sin\theta)$ and $\mathbf{n}_\perp = (-\sin\theta, \cos\theta)$ are the unit vectors along and perpendicular to the gradient direction [21], [54].

The basic snake model entails calibration of several parameters, has higher computational complexity, and cannot detect multiple objects in the same topology. However, it offers a better understanding of its interaction with MSOT images and perfectly suits the two compartment model of segmentation we are currently pursuing.

### Model-based reconstruction:

The IMMI algorithm [30] represents an exact numerical model-based reconstruction method and offers quantitative results by taking into account the various experimental imperfections, preserving the low frequency information and reducing image artifacts. In this method, the difference between the measured pressure at a set of locations and instants (expressed in a vector form as $\mathbf{p}$), and the equivalent theoretical pressure is iteratively minimized using least square minimization techniques. The optical absorption at the pixels of the ROI, expressed as vector form $\mathbf{F}$, is calculated as follows

$$\mathbf{F} = \text{argmin}_{\mathbf{f}} \parallel \mathbf{Af} - \mathbf{p} \parallel^2 + \lambda^2 \parallel \mathbf{Lf} \parallel^2 \qquad ..(15)$$

where $\mathbf{A}$ is the linear operator (or model matrix) mapping the optical absorption to the acoustic pressure. We use standard Tikhonov regularization to minimize the high-frequency noise and reduce the effects on limited view in the inversion process. The matrix $\mathbf{L}$ represents a high-pass filter operation. For all practical purposes we use a Butterworth band-pass filter between 0.1 -7 MHz to filter the acquired signal.